\documentstyle[12pt]{article}
\evensidemargin\oddsidemargin
\begin{document}
\count0 = 1
\begin{titlepage}
\vspace {20mm}
\begin{center}
\ QUANTUM REFERENCE FRAMES,\\
\smallskip
\ TIME AND MEASUREMENTS  \\
\vspace{6mm}

\bf{S.N. MAYBUROV}
\vspace{6mm}

\small{Lebedev Institute of Physics}\\
\small{Leninsky pr. 53, Moscow Russia, 117924}\\

\vspace{15mm}
\small{\bf Abstract}
\end {center}
\vspace{3mm}

\small{
 We argue that correct account of the quantum properties of macroscopic
 objects which form reference frames (RF) demand
 the change of the standard space-time picture accepted in
 Quantum Mechanics. The presence of RF free quantum motion in the form of
wave packet smearing results in formal nonapplicability of
Galilean or Lorentz space-time transformations in this case.  For the
 description of the particles states transformations
  between different quantum RF
 the special quantum space-time transformations are formulated.
 Consequently it results in corrections to Schrodinger or
  Klein- Gordon equations
 which depends on the RF mass. RF proper time becomes the operator
 depending of momentums spread in RF wave packet ,from the point of
 view of other observer. 
 The experiments with macroscopic coherent states are proposed
 in which this effects can be tested.} 

\vspace{14mm}
\vspace {20mm}
\vspace{28mm}
\small {---------}\\
\small {  * E-mail  Mayburov@sgi.lpi.msk.su}
\end{titlepage}
\begin{sloppypar}
\section{Introduction}

In a modern Quantum Mechanics (QM) the particles states and other objects
 evolve in Minkowski space-time regarded as independently existing
 entity. Alternatively it's
 the main method of the description of the surrounding  world from the point
 of  view of particular observer. In Classical Physics it corresponds to 
 introduction of space coordinate axes associated with particular    
 reference frame(RF) which is supposed to be some solid macroscopic object
 or the system of them. Despite that in QM
 the behavior of physical objects can be strikingly nonclassical it's
 tacitly assumed describing RF properties in QM
 that any RF evolution is always exactly classical. 
 Consequently all RF coordinate transformations in QM are supposed to be
 identical to classical - Galilean or Lorentzian ones.
 In our paper we argue that this assumption is in general incorrect and
 quantum features of RF should result in a special quantum space-time
 transformations. We must stress that this results are obtained without
introducing new axioms or hypothesis ,but staying in the framework of
 standard QM. It seems to us that current discussions of Quantum space-time
should include the correct definition of quantum RF (Doplicher,1995). 
The first consistent formulation of quantum RF problems was
given in (Aharonov,1984). Their algebraic and group properties were 
studied by Toller (Toller,1997).
 The importance of RF quantum properties was noticed in 
Quantum Gravity study of distributed - dust or fluid RF 
(Rovelli,1991). Yet the detailed analysis 
  of  Quantum Measurements problems concerned with 
quantum RF and quantum observers and the properties of
their proper time wasn't performed up to now.

In this paper only one quantum  RF effect out of many possible ones
 will be analyzed in detail. In fact it's the consequence of the 
 existence of the wave packet for any free macroscopic object,regarded as RF
 which width gradually enlarges with the time. Despite that this effect
 scale in the standard laboratory conditions is quite small ,it influences
on the coordinate measurements in principle, and we can expect
it can have important meaning both for Cosmology and for small 
 distance Physics.  It'll be shown that in general
  quantum RF transformations corresponds to the additional quantum
symmetry.  We describe briefly the special experiment which can test this
 conclusions  using modern experimental technic. Part of our results
were published earlier in (Mayburov,1997).   

 The formulated problem is closely related
 to the  macroscopic quantum
 coherence topics  which embrace different observations of the superpositions
 of the macroscopic objects states. The recent studies
 have shown that for low dissipation superconductor systems
the superpositions of the macroscopic states can be observed
 (Legett,1980).  

Our paper is organised as follows : in the rest of this chapter we remind
of the QM wave packet properties and formulate our model premises.
In a chapter 2 the new canonical formalism of quantum RF states and their
transformations  developed
and the analog of Schrodinger equation for finite mass observer is obtained.
 The possible experimental tests and their description by the 
Decoherence Model are considered in chapter 3. The
relativistic equations for quantum RF
and the resulting quantum space-time properties are regarded in 
chapter 4 . In a final chapter
the obtained results and their interpretation are discussed.

 Note that
 in QM framework the system defined as RF presumably should be able to measure 
the observables of quantum states i.e. to be observer i.e. to include
measuring devices -detectors.
 At first sight it seems this problem can be 
solved only when the detailed microscopic model of state vector collapse
 will be developed. Despite the multiple proposals up to now well established
theory of collapse which answer all difficult questions is absent
 (Mayburov,1995).
  Alternatively we'll show that our problem premises doesn't connected
 directly with the state vector collapse mechanism . To demonstrate it
clearly we'll perform in the framework of Decoherence model the calculation
of the particular example of such measurement. As the result 
in place of its detailed description we can use two simple assumptions
about the observer system properties , which are in the same time rather weak.
  The first one is that  RF consists of 
 finite number of atoms (usually rigidly connected)  and have the finite
 mass. We don't consider in our study the influence of detector recoil
 effects on the measurements results which can be made arbitrarily
small (Aharonov,1984).  

  So in this paper we'll assume that
observer system (OS) or RF
 is the ensemble of particle detectors , meters and
 recording
devices which perform the coordinate or other measurements on quantum
 objects. As the realistic example we can regard the photoemulsion plate or
the diamond crystal which can measure microparticle position relative to its
c.m. and simultaneously record it.
 We'll regard measuring device to be in a
 pure state as usually done in Measurement Theory.

 The experiments with the atomic
 and molecular beams confirm
 that the complex quantum system can be obtained in the delocalized state
without change of their internal properties.
  We'll consider the hypothetical
situation when one free observer $F^1$ is described by some other 
macroscopic observer as a pure quantum state with large
uncertainty of centre of
mass coordinate $R_c=\sum m_i*r_i/M$.
The question which we had in mind preparing this work was : if the observer
 can be in such
delocalised quantum state what will he see looking at the objects of
 our macroscopic world ?

 It's well known that the solution of Shrodinger equation for
  any free quantum system in a pure state consisting of $N$
constituents can be presented as the
\begin {equation} 
  \Psi(\vec{r}_1,...,\vec{r}_n,t)=
\sum c_l\Phi^c_l(\vec{R}_c,t)*\phi_l(\vec{r}_{i,j},t)
\end {equation}
where   $\vec{r}_{i,j}=\vec{r}_i-\vec{r}_j$  are  the relative or 'internal'
 coordinates of constituents
 (Schiff,1955). Here $\Phi^c_l$ describes the c.m. motion of the system.
 It demonstrates that in QM framework the functioning  and evolution of the
 system in the absence of the external fields is separated into the external
evolution as whole of the pointlike particle M and the internal evolution
 completely defined by $\phi_i(\vec{r}_{ij},t)$ if the constituents
 interaction
depends only on $\vec{r}_{ij}$ as usually have place. So the internal evolution is
 independent of whether the  system is localized
 in the standard macroscopic 'absolute' reference frame (ARF)
 or not. Relativistic QM and Field Theory studies show that the
 factorization of c.m.
 and relative motion holds true even for nonpotential forces and 
 variable $N$ in the secondarily quantized systems (Schweber,1961).
 Moreover this factorization expected to be correct for nonrelativistic
 systems 
 where binding energy is much less then its mass $m_1$, which is
 characteristic for the most of real detectors. For our problem
 it's enough to assume that
   the factorization of the c.m. motion holds for the observer system
  only in the time interval $T$
 from the system preparation procedure , until the act of measurement starts
,i.e. when the measured particle collides with it. More exactly our second and
 last assumption about observer properties is    
that  during period $T$ its state is described by the wave function
generalizing  (1) of the form 
$$
 \Psi(R_c,q,t)=\sum c_l\Phi^c_l(R_c,t)*\phi_l(q,t)
$$
 where $q$ denote all internal detector degrees of
freedom which evolve during $T$ according to Schrodinger equation
(or some field equation). To
 simplify our calculations
we'll take below all $c_l=0$ except $c_1$ which wouldn't influence our final
 results.
  
   Any free object
  which was initially localized in the wave packet $\Phi(R,t_0)$
 after it will gradually smear in space 
,which rate for the gaussian packet with initial dispersion $a_0$ described
as (Schiff,1955) :
\begin {equation}    
a(t)=(\int(R-\bar{R})^2|\Phi(R,t)|^2d^3R)^{1/2}
 =a_0(1+\frac{t^2}{m^2a_0^4})^{1/2} 
\end {equation}
(Plank constant $\hbar$ and $c$ in our calculations is equal $1$ ). 
The common conclusion is that to observe experimentally 
measurable smearing of macroscopic object
demands too large time , but we'll show that for some mesascopic experiments
it can be reasonably small to be tested in the laboratory conditions.  
In our work it's  permitted  ad hoc preparation of any initial
 state vector described by the smooth
function $\Psi(r_i,t_0)$ in agreement with QM postulates.

 We don't consider in our study the influence of detector recoil
 effects on the measurements results which can be made arbitrarily
small (Aharonov,1984). During this study we 
assume that RF and OS are always identical entities, but we'll discuss in
a final chapter their
possible distinctions which in fact can weaken demands to RF
formulated in this chapter.

\section{Q-transformations Formalism}

To explain the meaning of quantum RF transformations in QM 
we consider gedankenexperiment with the wave packet of observer system
 $F^1$ . Assume that
 $F^1$ of mass $m_1$ is suspended in a vacuum chamber which is regarded
as the classical absolute RF (ARF) with the mass $m_A\rightarrow\infty$
(gravitation field is absent).
 For the simplicity  $F^1$ c.m. wave packet
  is supposed to  smear  significantly only along X axe
described by $\psi_1(x)$. 
$$
  \psi_1(r_1,t)=\psi_1(x_1,t)\delta(y_1)\delta(z_1) 
$$
.
 The measured particle $n$
  with mass $m_n=m_2$ belongs to a very narrow beam
 , so that its wave function $\psi_n(x_n)$ can be approximated
 by the delta-function $\delta(x_n-x_s)$ . $F^1$ includes
particle detector $D_0$ which can measure the distance between the particle
and $F^1$ c.m. . 
 All states are taken at 
the fixed time $t=0$ when $n$ collides with $F^1$ and due to it  $t$ dependence
in $\psi$ arguments omitted. 

 Due to the factorization of $F^1$ and $n$ states
  $n$  wave function in $F^1$ $\psi'$ can be extracted from the
$F^1+n$  system wave function by the canonical transformation
\begin {equation}  
  \psi(x_n,x_1)= \psi_1(x_1)\psi_n(x_n)=\Phi_c(X_c)\psi'_n(x_n-x_1)=
 \psi_1(\frac{m_1x_1+m_nx_n}{m_1+m_n})\psi_1(x_n-x_1-x_s)
\end {equation}
Function $\Phi$ describes the state of this system as the whole and
can't be found by no measurement on $n$ in $F^1$. 
Parameters of $\psi'_n$  in principle can be defined by the measurement
of $n$ observables in $F^1$,which details wouldn't be discussed here.
 In this example $\psi'_n$ 
coincides with $\psi_1$ for $n$ being localized in ARF. 
 It assumes that
if in ARF $F^1$ wave function have the average
$x$ dispersion $a_O$ then from the 'point of view' of
 observer $F^1$ any object localized in ARF is
smeared with the same RMS $a_O$. 

Considering the collapse in this two RF 
 we note that
 $F^1$ and ARF observers will treat the same event unambiguously
as  $n$ detection  (or it flight through $D_0$). In
observer reference frame $F^1$  it
reveals itself by the detection and amplification process in $D_o$
 initiated by $n$ absorption. For ARF
the collapse results from the nonobservation of neutron in a due time
 - so called negative result experiment. So 
the signal in $F^1$ will have  the same relative probability as in
ARF. Such kind of the measurement means obviously the reduction of $\psi_1$
in ARF. Because of it proceeding further we'll assume 
always that all our considerations  are performed for the
 quantum ensemble of observers $F^1$
 without additional referring to it. It means that each event is resulted 
 from the interaction between the 'fresh' observer and also the particle
 ,prepared both in the specified quantum
 states ,alike the particle alone in the standard experiment.
 As we 
have no reason to assume that the transition from ARF to $F^1$ which we'll 
call Q-transformation can transfer 
pure states to mixed ones we must conclude that this distribution
is defined by neutron wave function in $F^1$. 
 It means that the result of measurement in $F^1$ is also
described by QM Reduction postulate, i.e. that initial state during
 the measurement by RF detector evolve into the mixture of
the measured observable eigenstates.

After this qualitative example we'll regard
the general situation for the system $S_N$ of $N$ objects $B_i$  which
 include $N_g$ pointlike 'particles' $G_i$ and $N_f$ frames $F^i$,
which in principle can have also some internal
 degrees of freedom described by (1).

For the start we'll assume that RF and particles coordinates observables
$\vec{r}_i$ are given
 in absolute (classical) ARF which is characterized by infinite
mass $m_A$ and coordinate $\vec{r_A}=\vec{r}_0=0$ and $\vec{p}_A=\vec{p}_0=0$
 in ARF ,denoted also as $F^0$ , but nonzero in some other ARF'.
 At the later stage we can abandon this notion and 
consider only relations between quantum RF and observables defined in them.
We should find two transformation operators - from ARF to quantum RF
,and between two quantum RF, but we'll show that in most general 
approach  they coincide .
We'll start from the former case and 
use  Jacoby canonical
coordinates  ${\vec{u}_j^l}$ ( $l=1,N_f$ denote corresponding RF)   
and conjugated momentums $\vec{\pi}^1_i$:
\begin{eqnarray}
  \vec{u}^1_1=\vec{r}_N-\vec{r}_{N-1} , \quad
 ,\vec{u}^1_{i}=\vec{r}^s_{N-i+1}-\vec{r}_{N-i} ,\nonumber\\
  \vec{u}^1_{N-1}=\vec{r}^s_{N-1}-\vec{r}_1 , \quad
 \vec{u}^1_N=\vec{R}_{cm}-\vec{r}_A \quad \label{B1}
\end{eqnarray}
where
$$
\vec{r}^s_i=\frac{\sum\limits^{N}_{j=i}m_j\vec{r}_j}{M^i_{N}} ,
$$
  $M^i_n=\sum\limits^{n}_{j=i}m_j$ (if upper index $i$ is omitted,
it assumed that $i=1$; also the vector sign omitted, where its use is obvious).
 The 'quantum' set $u^k$ can be obtained  from $u^1$ 
 changing cyclically $F^k$ and $F^1$ 
  in the $\vec{r}^i$  coordinates array so that
 $r'_1=r_k$, $r'_k=r_{k-1},...,r'_2=r_1$, $r'^s_i$ is defined by the
 above formula  in which $r'_j$ is substituted,
 and as the result ,$u^k_{N-1}=r'^s_{2}-r_k$ ,etc.
For  $j\neq1$, $\vec{r}_j-\vec{r}_1$ is the linear sum of several coordinates
 $\vec{u}^1_i$ ,so
 they don't commute ,due to quantum movement of $F^1$. 
 Conjugated to $u^1_i$ $(i=1,N)$ canonical momentums are :
$$
\begin {array} {c}
\vec{\pi}^1_1=\mu'_1(\frac{\vec{p}_N}{m_N}-\frac{\vec{p}_{N-1}}{m_{N-1}})\\
\vec{\pi}^1_i=
\mu'_i(\frac{\vec{p}^s_{N-i+1}}{M^{N-i+1}_N}-\frac{\vec{p}_{N-i}}{m_{N-i}})
\end {array}
$$
where $\vec{p}^s_i=\sum\limits^{N}_{j=i}\vec{p}_j$
 The transformed free Hamiltonian of the system objects motion is:
\begin {equation} 
 \hat{H}_c=\frac{(\vec{\pi}^{1}_N)^2}{2M_N}+
\sum\limits_{j=1}^{N-1}\frac{(\vec{\pi}^{1}_i)^2}{2\mu'_i}
\end {equation}
,where reduced mass
 $ \mu'^{-1}_i=(M^{N-i+1}_N)^{-1}+m_{N-i}^{-1} $ .

To find the transformation between 2 quantum RF  we start from
the simplest case $N_f=2,N_g=0$. This is just the space reflection
of  $F^1$ coordinate $u^2_1=-u^1_1$ performed by the parity operator
 $\hat{P}_1$.  This discrete transformation
is presented in the indirect form in all other cases.
 The next case $N_f=2,N_g=1$ is just
the two $u^1$ coordinates linear transformation exchanging $r_2,r_1$ :
$$
  u^2_{1,2}=\hat{U}_{2,1}u^1\hat{U}^{+}_{2,1}=a_{1,2}u^1_1+b_{1,2}u^1_2
$$
The unitary operator $\hat{U}_{2,1}$  in general can be decomposed as 
$\hat{U}=\hat{C_2}\hat{R}\hat{C_1}$ ,where $\hat{C_{1,2}}$
is conformal transformation of the kind $u'^1_i=c_iu^1_i$
,such that $c_1c_2=1$ 
$$
   c_1=[\frac{M_3m_3m_2}{m_1(m_2+m_1)^2}]^{\frac{1}{4}}
$$
 $R$ is the rotation on $u'^1_{1,2}$ intermediate
coordinates hypersurface on the angle :
$$
\beta=\arccos[\frac{m_2m_1}{(m_3+m_2)(m_1+m_3)}]^\frac{1}{2}
$$
Then $\hat{C}_2$ results in $u^2_i=c'_iu'^1_i$.
 For the general case $N>3$ it's possible nonethereless
to decompose the transformation from $F^j$ to $F^k$ 
as the product of such bilinear operators. Really if to denote
as $\hat{S}_{i+1,i}$ the operator exchanging $F_i,F_{i+1}$  in $u^1$ set,
which changes in fact only
$u^1_i,u^1_{i+1}$ pair ,values of other $u^1_j$ as easily seen
conserved under this exchange; and  $\hat{U}_{2,1}=\hat{S}_{2,1}$ obviously. 
Then the transformation operator
 from $F^1$ to $F^k$ is :
$$
   \hat{U}_{k,1}=\hat{U}_{2,1}\hat{S}_{3,2}...\hat{S}_{k,k-1}
$$  
It follows immediately that the transformation from $F^j$ to $F^k$ is
 $\hat{U}_{j,k}=\hat{U}_{k,1}\hat{U}^{-1}_{j,1}$.

Now we must find the transformation operator from the classical ARF
 to $F^1$, where
ARF 'classical' set $\vec{u}^A_i=\vec{r}_{N-i+1}-\vec{r}_A$.
 Here we must formally
consider each object as RF to perform the intermediate transformations and
include ARF  in $F^i$ array as $F^{N+1}$ ,so $N'_f=N'=N+1$.
Note that the set $u^1$ can be rewritten as the Jacoby  'quantum' set for
ARF if we formally add to it the $\vec{u}^{N+1}_{N+1}=\vec{r}_A-\vec{r}_E$ 
,where $E$ is some
other classical RF.
  Then it's easy
to observe that acting by $\hat{P}_1$ operator ,resulting in $u'^A_N=-u^A_N$
gives $u^1$ 'quantum' set of (\ref{B1}) for $m_{N+1}\rightarrow\infty$.
 We must add to it
formally  $u^1_{N+1}=u^{N+1}_{N+1}$ and then the operator in question
 is equal to $\hat{U}_{A,1}=\hat{U}_{N+1,1}\hat{P}_1$ for infinite $m_{N+1}$.

 Note that even settling
$\vec{r_1}=0$, $\vec{p_1}=0$ for $F^1$ we must account them as the operators in
commutation relations ,as was stressed in (Dirac,1956). Neglecting
it result in so called 'Quantum Frames Paradox' (Aharonov,1984) ,which have no
independent significance.

 This problem become more intricate if we want to account 
the possible quantum rotation of our RF relative to ARF.
  We'll  consider here only stationary rotational states 
 for 2-dimensional rotations. If $F^1$ is the solid object its orientation  
 relative to ARF can be extracted from the relative (internal) coordinates 
 of $F^1$ constituents (atoms). For the simplicity we assume that
 $F^1$ have the dipole structure and all its mass concentrated around
2 points $\vec{r}_{a1},\vec{r}_{b1}$ so that this relative coordinate is
$F^1$ independent degree of freedom   
  $\vec{r}_{a1}-\vec{r}_{b1}$ or in polar coordinates $r^d_{1},\theta_1$.
Note that $r^d_1$ is observable which eigenvalue defined by $F^1$ 
constituents interaction. 
 Thus after
 performing transformation $\hat{U}_{A,1}$ to $F^1$ c.m. we'll rotate all the 
objects (including ARF) around it on the uncertain angle $\theta_1$
,so the complete transformation is
 $\hat{U}^T_{A,1}=\hat{U}^R_{A,1}\hat{U}_{A,1}$.
In its turn this rotation operator can be decomposed as
$\hat{U}^R_{A,1}=\hat{U}^c_{A,1}\hat{U}^d_{A,1}$, representing the
rotation of objects c.m coordinates $u^1_i$ and $F^i$ constituents coordinates.
 This rotation introduces the quantum uncertainty
of ARF orientation relative to $F^1$.  
The rotation of $F^1$ is performed by the operator $\hat{U}^d_{A,1}$, 
which action settles $\theta_1$ to zero ,and introduce in place of it
the new observable $\theta^r_1$ which corresponds to ARF  angle in $F^1$. 

 The objects c.m. coordinates  transformation operator is: 
 $$
\hat{U}^c_{A,1}=e^{-i\theta_1L_z} ,\quad
L_z=\sum\limits^{N}_{i=1}l_{zi} ,\quad 
 l_{zi}=-id/d\alpha_i
$$
 where $\alpha_i$ is the polar angle coordinate of $\vec{u}^1_i$ .
It results in :
\begin {eqnarray}
  u^{1r}_{xi}=u^1_{xi}\cos\theta_1+u^1_{yi}\sin\theta_1 \label{B3}\\
  u^{1r}_{yi}=-u^1_{xi}\sin\theta_1+u^1_{yi}\cos\theta_1 \nonumber
\end {eqnarray}
So the new polar angle is $\alpha^r_i=\alpha_i-\theta_1$.
The transformation of the external objects momentums is analogous :
$$
\begin {array}{c}
 \pi^{1r}_{xi}=\pi_{xi}\cos\theta_1+\pi_{yi}\sin\theta_1\\
 \pi^{1,r}_{yi}=-\pi_{xi}\sin\theta_1+\pi_{yi}\cos\theta_1
\end {array}
$$
As easily seen Hamiltonian $\hat{H}_c$ of (5)  is invariant under this
 transformation.
We must account also rotation of any $F^i$ internal degrees of freedom
 described by formula (1).
  Assuming that any $F^i$ have analogous to $F^1$ dipole form
 their 'internal' Hamiltonian in ARF is :
\begin {equation} 
\label {B9} 
 \hat{H}_i=\sum\limits_{j=1}^{N_f}[\frac{1}{m^d_jr^{d2}_{j}}
\frac{\partial^2}{\partial\theta_j^2}+V_j+\hat{H}^f_{j}]
\end {equation}
where $m^d_j$ is the effective mass of the rotational moment which
for the dipole is equal to its reduced mass, $\hat{H}^f_{j}$ is the
part of $F^j$ constituents free Hamiltonian of their relative motion
which is rotationally invariant.
$V_j$ is the potential  of $F^j$ constituents interaction and is
obviously invariant of $F^j$ rotation.

 If to denote $\hat{P}^d_1$ parity operator for $\theta_1$
 $l^d_{j}=-i\frac{\partial}{\partial\theta_j}$ then
\begin {eqnarray*}
\hat{U}^d_{A,1}=\hat{P}^d_1exp(i\theta_1 L_d) ,
\quad L_d=\sum\limits_{i=2}^{N_f}l^d_{i}\\
\theta^r_j=\theta_j-\theta_1 ,\quad l'^d_{j}=l^d_{j} \quad j\neq1\\
\theta^r_1=\theta^r_A=-\theta_1 ,
 \quad l'^d_1=l^d_{A}=-l^d_{1}-L_d
\end {eqnarray*}
The new coordinates can be interpreted as corresponding to 
$F^1$ dipole rest frame , where its own angle $\theta'_1$ is fixed to zero
but ARF angle in $F^1$  becomes uncertain and formally ARF acquires
the orbital momentum $l^d_{A}$.

The rotational part of Hamiltonian $\hat{H}_i$ in this rest frame
expressed through the new coordinates is :
$$
 \hat{H}^r_i=\sum\limits_{j=2}^{N_f}\frac{1}{m^d_{j}r^{d2}_j}l'^{d2}_{j}+
\frac{(l'^d_{1}+\sum\limits_{j=2}^{N_f}l'^d_{j})^2}{m^d_1 r^{d2}_1}
$$
So we get the conclusive and noncontroversial description of
$G^i$ and $F^i$ evolution in $F^1$ defined by Hamiltonian
 $\hat{H}_c+\hat{H}_i$ .
 Yet this transformation can result in the change of the objects $G^i$
 wave functions $\Psi^1$ in $F^1$ which will depend on $F^1$ orbital
 momentum
which should be accounted performing the initial functions  transformation. 

Analogous considerations permit to find rotational transformation
$\hat{U}^R_{i,1}$ from $F^1$ to $F^i$. We note that it just the
additional rotation
of all the objects on the  angle $\theta^r_i=\theta_i-\theta_1$. 
So the form of $\hat{U}^c$ part of rotational operator is unchanged and
$$
\hat{U}^d_{i,1}=\hat{P}'^d_ie^{-i\sum\limits_{j=1}^{N_f}l'^d_j\theta^r_i}
e^{il'^d_i\theta^r_i}
$$ 
,where $\hat{P'}^d_i$ is parity operator for $\theta^r_i$.

For $d=3$ the mathematical calculations are analogous ,but more tedious,
if to remind that any rotation in space can be decomposed as
three consequent rotations in the specified orthogonal planes. 
  So we omit the calculations for $d=3$ here, and just explain what
the new features appears. To describe
  this rotation $F^1$ should have the necessary structure
, the simplest of which is the rigid triangle $abc$ with constituents masses
concentrated in its vertexes. Then $Z'$ axe can be chosen to be orthogonal 
to the triangle plane and $X'$ directed along $ab$ side. Then the transformation
which aligns ARF and $F^1$ axes can be performed  rotating consequently
ARF around $X',Y',Z'$ on the uncertain angles $\theta_x,\theta_y
, \theta_z$. Each of three operators performing it is the analog of
 $\hat{U}^R_{A,1}$ described above.
  
Now we regard the evolution equation in quantum RF taking the time $t$ 
as universal parameter independent of RF. Note that standard 
Schrodinger equation assumes mutely that observer for which
the wave function is defined have infinite mass. For such observer ARF
we define the system wave function $\psi_s(\vec{r}_i,t)$
 in the standard picture. 
  It satisfy to free Schrodinger equation
for N objects and with some initial conditions can be factorised into
 $\Phi_c(\vec{R}_c,t)\psi'_s(\vec{u}^j,t)$ as we discussed in chap. 1.
Then the resulting equation for the relative motion of N objects of $S_N$
can be obtained ,if we remove system c.m. motion from $\hat{H}_c$ of (5).
 The resulting equation for $\psi'_s$ -  
 wave function in $u^1$ coordinates is :
\begin {equation} 
-\sum_{j=1}^{N-1}\frac{1}{2\mu'_j}\frac{\partial^2}{\partial\vec{u}^{12}_j}
 \psi'_s(u^1,t)
   =i\frac{d\psi'_s}{dt}(u^1,t)
\end {equation}
It is reduced to Schrodinger equation ,if $m_1\rightarrow\infty$ and 
is analogous to the equation for relative $e-p$ motion in Hydrogen atom
(Schiff,1955). Yet in distinction in $F^1$ rest frame ARF coordinate
$\vec{r}_0-\vec{r}_1$ becomes uncertain and to calculate its evolution
through the coordinate $u^1_N$ in this equation
we must use complete Hamiltonian $\hat{H_c}$.  
Then it's easy to understand that if $m_0$ isn't infinite
 ,but just large we must change  $M_N$ to $\mu`_N$ in the first 
$\hat{H}_c$ member in (5). We must notice the different form 
of $\vec{u}^1_N,\vec{\pi}^1_N$ dependence 
on $\vec{r}_1$ in comparisons with other
$\vec{u}^1_i,\vec{\pi}^1_i$. This can be regarded as ARF doesn`t belong to the
studied system $S_N$ ,but is the `outsider` object. 
This form of Jacoby coordinates can be extended in an 
obvious way for the description of the state of
any other object of the universe.
We conclude that equation (8) ,which can be modified if necessary to
describe the evolution of 
`outsider` objects in $F^1$ is the correct evolution
equation which depends on observer $F^1$ mass.

\section {Measurements aspects} 

Now we'll discuss briefly the feasibility of the experiments
with RF wave packets.   
Its principal scheme is analogous to described above gedankenexperiment
 where   in  the  vacuum chamber 
 the solid state detector-recorder initially rigidly fixed
  is released  and suspended at $t=-T_L$. After some time period 
 the detector performs the coordinate measurement of the particle $n$ 
 from the very narrow beam, which permit to find the detector quantum
 displacement.
 As the detector-recorder system
in fact can be used any detector with the memory like the  photoemulsion which
have coordinate accuracy of the order .1 micron.
Especially attractive  seems to be the plastic or crystal track detectors 
which under the electron microscopic scanning can in principle define  the
position of dislocation induced by particle track with the 
accuracy up to several interatomic distances. The same order will have
the initial packet smearing $a_0$, because it's defined by the surface
effects between the detector and fixator surfaces which extended to 
the  interatomic scale.      
 If we suppose the mass of the detector
to be $10^{-10}$ gram (the mass of emulsion grain which acts as the
elementary individual detector) and 
 $a_0$ value $10^{-2} mk$ we get 
the average centre mass deflection
of the order .1 mk for the exposition time $10^6$ sec i.e. about
one week.
Despite that the performance of such experiments will be technically
extremely difficult it's important nonethereless that 
 no principal prohibitions  for them exist.    

Now we'll consider the analogous experiment  description in the framework of
Decoherence model (DM) of quantum measurements (Zurek,1982).
 It's reasonable to take that the result of
the particle-grain interaction is roughly dichotomic : or $n$ passes by
without changing  grain D initial state $|U_{-}\rangle$ ,or passes 
through and darken it corresponding 
 to $|U_+\rangle$, i.e. to assume that D have the single
dichotomic degree of freedom (DF). 
 If the  consequent interaction of D with
 the environment E results into the state collapse as DM assumes, 
then the crude measurement of the distance $\delta r$ between
 $n$ and D c.m. occurs. Let's rewrite $n$ state vector in the discrete form 
$$
  |\psi^n_0\rangle=a|+\rangle+b|-\rangle
$$
where $|+,-\rangle$ are state components inside and outside $r_G$ 
vicinity.  We'll assume that $n$-D interaction ,which Hamiltonian is given 
below turns in at $t=-T$ and turns off at $t=0$ , so that $T$ roughly
corresponds to$n$ time of flight through the grain.
$$
   H^{n-D}=g_0(1+\sigma_n^z) \sigma^x_u=g_0|+ \rangle \langle+|
(|U_+\rangle \langle U_-|+|U_- \rangle \langle U_+|)
$$
We choose $T$ to be equal to :
$$
    T=\frac{\pi\hbar}{2g_0}
$$

 Then at $t>0$ D starts to interact
with E having very large number $N_E$ of DF. In the regarded set-up its
 realization means that the vacuum chamber is open and some gas
which molecules have single dichotomic DF $|^+_{-i}\rangle$
 filled it.
We assume that in  $n$-D and D-E Hamiltonians
 free parts for $n$,D  in the evolution equations can be neglected.
Their account doesn't change results principally ,but makes the calculations
more complicated (Bell,1975). D-E interaction Hamiltonian is additive
relative to E molecules, and the interaction between the molecules and their
 free motion  neglected:
$$
   H^{D-E}_i=\sum\limits_{i=1}^{N_E} g_i\sigma^y_u\sigma^y_i
$$
In Zurek model $g_i$ - E coupling constants and initial E molecules 
state vectors
${\alpha_i,\beta_i}$ values are distributed at random independently
 of each other ,so that $g_{min}<g_i<g_{max}$.
Solving Shrodinger equation we find $\Psi(t)$ from which the density matrix
$\rho(t)$
of $n$-D system can be obtained tracing over $E$ DF ,which assumedly 
are unobservable.
$$
 \rho(t)=|a|^2|s_+\rangle \langle s_+|+|b|^2|s_{-}\rangle \langle s_-|
+iz(t)ab^*|s_+\rangle \langle s_+|-iz^*(t)a^*b|s_{-}\rangle \langle s_{-}|
$$
where $|s^+_-\rangle=|^+_-\rangle|U^+_-\rangle$,
$$
 z(t)=\prod_{k=1}^{N_E}[\cos2g_kt+i(|\alpha_k|^2-|\beta_k|^2)\sin2g_kt]
$$

Zurek showed that for large $N$ $z(t)$ soon become very small and 
the corresponding difference between $\rho(t)$ and $\rho_M$ - matrix
 of completely mixed state also becomes very small. So  we conclude that
that the state of $n$-D system during its evolutions 
approximates to the collapsed mixed state,
 corresponding to the measurement of $n$-D relative distance.
 So it seems that the state evolution in this delocalized
 measurements mainly follows the collapse postulate.
 This results supports our point that it's correct to restrict ourself
to the two assumptions formulated in chap.1 in place of using
 any detailed collapse model.  

\section  {Relativistic Equations}

Obviously the most important aim for this model development
is to give relativistically covariant quantum RF description,
which will describe also the quantum time properties.
Here we'll argue that they are mainly 
analogous to the space coordinates properties found in the former chapter. 
We'll apply the same approach  concerned with the relativistic wave packets
of solid objects -regarded as quantum RF .  
 We'll suppose that this observer can be described
 as the zero-spin boson
,because normally all its constituents spins are roughly compensated.
For the simplicity we'll consider only 1-dimensional situations which
permits us to neglect the  quantum rotations.

In nonrelativistic mechanics time $t$ is universal and 'external' relative to 
any system and so is independent of any observer. In relativistic
case  each observer in principle has its own proper time $\tau$ 
measured by his clocks, which must be
presented in relativistic evolution equation in his RF .

We don't know yet the origin of the physical time , but for the 
modelling purposes we can associate it with the clock hands motion,
or more exactly with the measurement and recording of their current position
 by observer. 
This motion is stipulated by some irreversible processes
which are practically unstudied on quantum level and we'll consider below
the simple model which approximates such processes. In our opinion
there is a strong and deep analogy between irreversible wave function
collapse in the measurement and clock hands motion+measurement
 which can be regarded as
the system self-measurement (Horwitz,88).
 It was argued recently that physical time
can originate from some non-Hamiltonian dynamics,
because any Hamiltonian Dynamics automatically guarantees reversibility
 (Rovelly,90). 
Meanwhile without choosing one or other mechanism , it's possible
 to assume that as in the case of the position measurement this internal
 processes
 can be disentangled from the clocks c.m. motion. Then clocks $+$ observer
 i.e. RF $F^2$ wave packet evolution can be described by the
 relativistic equation 
  for their c.m. motion. In this packet
 different momentums and consequently velocities  relative to
external observer $F^1$ are presented. It makes impossible to connect
external time $\tau_1$ and $F^2$ proper time $\tau_2$ by any Lorentz
transformation, which corresponds to the unique definite
Lorentz factor ${\gamma}(\vec{v})$.

To illustrate the main idea we remind the well-known
situation with the relativistic lifetime dilation of unstable particles
or metastable atoms. Imagine that  the prepared beam of them 
is the superposition of two or more momentums eigenstates having
different Lorentz factors $\gamma_i$.
 Then detecting  their  decay  products  we'll
find the superposition of several lifetime exponents , resulting from the
fact that for each beam component Lorentz time boost has its own value.
If in some sense this unstable state can be regarded as elementary
 clock when their time rate for the external observer is defined by the
superposition of Lorentz boosts responding to this momentums.

From this arguments we can assume that the proper time of any quantum RF
 being the parameter in his rest frame simultaneously
 will be the operator for  other quantum RF.
If this is the case the proper time $\tau_2$ of $F^2$ in $F^1$ can be
 the parameter
depending operator, where parameter is $\tau_1$.
$$ 
    \hat{\tau}_2=\hat{F}(\tau_1)=
 \hat{B}_{12}\tau_1
$$
 ,where $\hat{B}_{12}$ can be called Lorentz boost operator,which 
can be the function of $F^1,F^2$ relative momentum. To define its form
it's necessary first to find Hamiltonian of free particle $G^2$ in $F^1$. 
Obviously in relativistic case Hamiltonian of relative motion
 of very heavy RF and light particle
  should approximate Klein-Gordon square root
 Hamiltonian, but in general it can differ from it (Schweber,1961).
The main idea  how to find it is the same as in nonrelativistic case
 : to separate the system c. m.
 motion and the relative motion of the system parts ,but  in  relativistic
case this is much more complicated problem (Coester,1965).
For the simplicity we'll consider first the evolution of
 RF $F^1$ and the particle $G^2$
which observables are defined in classic ARF.
 In  Classical Relativity the
objects relative motion is characterized by their invariant mass square $s^m$,
which is equal to system total energy in its c.m.s.,
equivalent of nonrelativisitic c.m. kinetic energy $E_k$.
 In our case it's equal to :
$$
s^m_{12}=(m_1^2+\vec{q}_{12}^2)^{\frac{1}{2}}+
(m_2^2+\vec{q}_{12}^2)^{\frac{1}{2}}
$$ 
,where $\vec{q}_{12}$ is  $G^2$  momentum in c.m.s. .
We  also define $G^2$ momentum in
$F^1$ rest frame which corresponds to Klein-Gordon momentum operator :
$$
  \vec{p}_{12}=\frac{s^m_{12}\vec{q}_{12}}{m_1}=\frac{E_1\vec{p}_2
-E_2\vec{p}_1}{m_1}  
$$
where $E_i,\vec{p}_i$ are total energies and momentums in ARF.
 We can expect that Hamiltonian of free particle in $F^1$ will correspond to 
4-th component of $\vec{p}_{12}$ 3-vector. Really
 if to transform
$F^1,G^2$ total momentum in c.m.s., which 4-th component is 
 $s^m_{12}$  to $F^1$ rest frame one obtains :  
$$
  E_1=[(s^{m}_{12})^2+\vec{p}^2_{12}]^\frac{1}{2}=
m_1+(m_2^2+\vec{p}^2_{12})^\frac{1}{2}
$$  
 Following this classical analysis in accordance with Correspondence
Principle
 we can  regard  $E_1$ as possible Hamiltonian $\hat{H}^1$
in $F^1$ rest frame
and the evolution equation for $G_2$ 
 for corresponding proper time $\tau_1$ is :
\begin {equation} 
\hat{H}^1\psi^1(\vec{p}_{12},\tau_1)
 =-i\frac{d\psi^1(\vec{p}_{12},\tau_1)}{d\tau_1}
\end {equation}
It's easy to note that $\hat{H}^1$ depends only of relative motion
observables and in particular can be rewrited as function of $\vec{q}_{12}$.
 This equation coincides with Klein-Gordon one, where it's possible to
consider  $m_1$ as arbitrary constant added to total energy. 
Consequently we can use in $F^1$ the same momentum eigenstates spectral 
decomposition (Schweber, 1961)

 Space coordinates in $F^1$ is difficult
 to define unambiguously, as usual in relativistic problems, so we choose
the general form of Newton-Wigner ansatz (Wigner,1986).
$$
 x_{12}=i\frac{d}{dp^x_{12}}+F_x(\vec{p}_{12})
$$
 
If we consider the evolution of RF $F^2$ in place of $G^2$ described by the
 same equation
,then its proper time operator
 $\tau_2$ in $F^1$ can be defined 
 from the correspondence with the classical Lorentz time boost
as :
$$ 
    \hat{\tau}_2=
 \hat{B}_{12}\tau_1=[(s^m_{12})^2-m_1^2-m_2^2]^{-1}m_1m_2\tau_1
$$
 Note that this form is completely symmetrical and the
same operator relates the time $\hat{\tau_1}$ in $F^1$ and $F^2$ proper time
- parameter $\tau_2$.
Despite this novelty no qualitatively new physical effects for
 the individual observer $F^2$ 
in addition to described in the previous chapters appears.
By himself (or itself ) $F^2$ can't find any consequences of time arrow
superpositions registrated by external $F^1$ ,
 for $F^2$ exists only unique proper time $\tau_2$ .
 The only new effect will be found
when $F^1$ and $F^2$ will compare their initially synchronized clocks.
If this experiment will be repeated identically several times
(to perform quantum ensemble) they find not only the standard
Lorentz proper times difference ,
 but also the statistical spread having quantum origin and
proportional to the time interval and $F^2$ momentum spread.

Analoguously to Classical Relativity average time boost depends on whether $F^1$ 
measures $F^2$ observables, as we considered or vice versa. To perform this 
measurement  we must have at least two synchronized
 objects $F^1_a$ and $F^1_b$ ,which make two $F^1$ and $F^2$ 
nonequivalent. 

If the number of objects $N>2$
we'll  use multilevel 'clasterization' formalism,
 which will be described here for 
the case $N=3$  (Coester,1965). In its framework
 Hamiltonian in $F^1$ and describing the two particles $G^2,G^3$ 
state evolution for  proper time $\tau_1$ is equal to :
$$
   \hat{H^1}=
   m_1+[(s^{m}_{23})^2+\vec{p}^2_{1,2+3}]^\frac{1}{2}
$$
,where $s^m_{23}$ is  two particles $G^2,G^3$ 
invariant mass given above. In this
formalism at first level we consider their relative motion defined by
$\vec{q}_{23}$. At second level we regard them as the single quasiparticle
 - claster with mass $s^m_{23}$. Then $\vec{p}_{1,2+3}$ is the cluster 
  momentum in $F^1$. So at any level we regard 
 the relative motion of two objects only. Despite that this Hamiltonian isn't 
 factorized between this two levels some of its feature is analogous to
 the nonrelativistic one of (5). We can extract small Hilbert space $H^1_2$
 ,which basis is $|\vec{p}_{12}\rangle$ 
 from the total space $H^1_s$ (Coester,1965). 
 In this small space the evolution of $m_2$ state $\psi'(\vec{p}_{12},\tau_1)$ 
 defined in $F^1$ can be
 described completely. The evolution of $m_3$ depends in its turn
 on $s^m_{12}$ and it means
  in fact one-way factorization of the claster momentums states. This
 procedure can be extended in the obvious inductive way to incorporate 
 an arbitrary number of the objects.


Now we'll regard the simple and crude model of the quantum clocks and RF
in which $F^1$ includes some ensemble (for example the crystal) of 
$\beta$-radioactive atoms. Their nucleus can radiate neutrino $\nu$
(together with the electron partner)
which due to its very small cross-section practically can't be reflected
 by any mirror  and
reabsorbed by this nucleus to restore the initial state. Then for our purposes
this decay can be regarded as the irreversible stochastic process.
Taking the trace over $\nu$ degrees of freedom,
the final nucleus state can be described by the density matrix of
mixed state $\rho_N(t)$
and the proper time of this clocks of $F^1$ can be defined as:
$$ 
  \tau_1=-T_d\ln(1-\frac{N_d}{N_0})  
$$
where $N_0$ is the initial number of this atoms $N_d$ - the number of decays,
$T_d$ is the nucleus lifetime .
It's easy to understand from the previous discussion how
 the superposition of Lorentz boosts can be apllied to such
  system state, if it has momentum spread .

  We consider in fact infrared 
limit for macroscopic object, so the role of negative energy states,which
is important for the standard relativistic problems must
be small. Despite the locality paradox  found for
the relativistic wave packets  we can expect this solutions to be valid
at least in the infrared limit and beyond , when the packet size is much 
larger then Compton length (Hegerfeldt,1976).

\section{\large{Concluding Remarks}}

  We've shown that the extrapolation of QM laws on the
macroscopic objects demands to change the approach to the 
space-time coordinate frames which was taken copiously from  Classical 
Physics. It seems that QM permits the existence of RF
 manifold, the transformations between which principally can't
be reduced to Galilean or Lorentz transformations.
This new global symmetry means that observer can't measure its own
spread in space, so as follows from Mach Principle it doesn't exist.
The physical meaning have only the spread of relative coordinates
of RF and some external object which can be measured by this RF or other
observer. 

Historically QM formulation started from defining the wave functions on
Euclidian 3-space $R^3$ wich constitute Hilbert space $H_s$.
 In the alternative approach
developed here we can regard $H_s$ as primordial states
 manifold. Introducing
particular Hamiltonian results in the relative assymmetry of $H_s$ vectors
 which permit
 to define $R^3$ as a spectrum of the continuous observable $\hat{\vec{r}}$
which eigenstates are
 $|\vec{r}_i>$. But as we've shown for several
 quantum objects one of which is RF
this definition
become ambiguous and have several alternative solutions defining $R^3$
on $H_s$. In the relativistic case the situation is more complicated, yet
as we've  shown it results in ambiguous Minkovsky space-time definition. 
    
At first sight Q-transformation will violate Locality principle,but
it's easy to see that it holds for each particular RF, despite that 
the coordinate point in
one RF doesn't transforms into the point in other RF. This is easy to see for
the nonrelativistic potential $V(r_2-r_1)$ ,but we can expect it true also
in relativistic Field Theory. 
 So we can suppose
the generalisation of locality principle for Q-transformations, which 
yet must be formulated in a closed form.  

In our work we demanded strictly that each RF must be quantum observer
i.e. to be able to measure state vector parameters. But we
should understand whether this ability is main property
 characterising RF ? In classical Physics this ability 
doesn't influence the system principal dynamical properties. In QM at first
sight we can't
claim it true or false finally because we don't have the established theory
 of collapse. 
 But it can be seen from our analysis that collapse is needed
in any RF only to measure the wave functions parameters at some $t$.
Alternatevely this parameters at any RF can be calculated given
the initial experimental conditions without performing the 
additional measurements.
It's quite reasonable to take that quantum states have objective meaning
and exist independently of 
their measurability by the particular observer,so this ability probably can't
 be decisive for this problem. It means that we can connect RF with the
system which doesn't include detectors ,which can weaken and simplify
our assumptions about RF. We can assume that more important for the 
object to be regarded as RF is the ability
to reproduce space and time points ordering and to record it
 ,as the  solid states like the crystalls and the atomic clocks can do.

\end{sloppypar}


\begin{thebibliography}{99}

\bibitem{Aha} Y.Aharonov,T.Kaufherr Phys. Rev. D30 ,368 (1984)

\bibitem {Bell} J.S.Bell Helv. Phys. Acta 48,93 (1975)

\bibitem {Dir} P.A.M.Dirac 'Lectures on Quantum Mechanics', (Heshiva
University ,N-Y,1964)

\bibitem{Dop} S.Doplicher et. al. Comm. Math. Phys. 172,187 (1995) 

\bibitem {Heg} G.C. Hegerfeldt Phys. Rev. Lett.54,2395 (1985)

\bibitem{Hor} L.Horwitz et. al. Found. Phys. 18,1159(1988)
 
\bibitem{Leg} A.J.Legett, Progr. Theor. Phys. Suppl. 69,80(1980)
,also R.J. Prance et. al.`Observation of Quantum jumps
in Squid Rings',Proc.
of 2nd conference of quantum communications and measurements.
(New-York, Plenum Press,1995).

\bibitem{May} S.N.Mayburov,  Int. Journ. Theor. Phys. 34,1587 (1995)
, see also W. D'Espagnat Found. Phys. 20,1157,(1990)

\bibitem{May2} S.N. Mayburov ,'Quantum Reference Frames', in 'Proc.
of  6th Quantum Qravity Seminar', Moscow, 1995 (W.S.,Singapore,1997)

\bibitem {Rov} C.Rovelli, Phys. Rev. D42 ,2638,(1990)

\bibitem {Rov2} C.Rovelli, Class. Quant. Grav. 8, 317,(1991)
 
\bibitem {Shw} S. Schweber 'An Introduction to Relativistic Quantum
 Field Theory' ,(New-York,Row,Peterson ,1961) 


\bibitem{Schiff} L.I.Schiff, 'Quantum Mechanics' (New-York, Macgraw-Hill,1955) 

\bibitem {Tol} M. Toller , Subm. to Nuov. Cim. B (1997)

\bibitem {Von} J.Von Neuman 'Mathematical Foundations of Quantum
 Mechanics' (Prinston ,1961) 

\bibitem{Wigner} E.Wigner Int. Journ. Theor. Phys. 25,467 (1986)
,with T.Newton Rev.Mod. Phys. 21,400,(1949)

\bibitem{Coe} F.Coester Helv. Phys. Acta 38,7 (1965)

\bibitem {Zur} W.Zurek Phys.Rev. D26,1862,(1982)
\end{thebibliography}
\end{document}